\documentclass[aps,article]{revtex4}%
\usepackage{amsfonts}
\usepackage{amsmath}
\usepackage{amssymb}
\usepackage{graphicx}%
\begin{document}
\title{A chiral $\overline{q}\overline{q}qq$ nonet? }
\author{Mauro Napsuciale and Sim\'{o}n Rodr\'{\i}guez }
\affiliation{37150, Le\'{o}n, Guanajuato, M\'{e}xico.}

\begin{abstract}
We point out that meson spectrum indicates the existence of a degenerate
chiral nonet in the energy region around $1.4\ GeV$ with a slightly inverted
spectrum with respect to a $\overline{q}q$ nonet. Based on this observation
and the approximately linear rising of the mass of a hadron with the number of
constituent quarks we conjecture the existence of a tetraquark chiral nonet in
this energy region with chiral symmetry implemented directly. We realize this
idea in a chiral model and take into account the mixing \ of the tetraquark
chiral nonet with a conventional $\overline{q}q$ nonet. We find that the mass
spectrum of mesons below $1.5\ GeV$ is consistent with this picture. In
general, pseudoscalar states arise as mainly $\overline{q}q$ states but scalar
states turn out to be strong admixtures of $\overline{q}q$ and tetraquark states.

\end{abstract}
\maketitle


\bigskip

The understanding of the puzzling scalar mesons is of primary interest since
these mesons have the same quantum numbers as the QCD vacuum and their
properties are expected to be strongly influenced by the latter. The hope is
that a better understanding of the properties of scalar mesons yield some
information on the structure of the QCD vacuum.

The quark structure of mesons is usually inferred from the accuracy of quark
model predictions for their properties or the interpretation of their decay
products according to a flavor structure. This structure is mapped onto the
transformation properties of quark composites under the $SU(3)_{F}$ group. In
the case of the scalars, the cuasi-degeneracy of the firstly discovered
$f_{0}(980)$ and $a_{0}(980)$ mesons suggesting an $\omega-\rho$ like system
was in contradiction with the strong coupling of the former to $K\overline{K}$
system. It was also soon realized that naive quark model calculations fail to
describe the small two photon decay widths of the $f_{0}(980)$ and
$a_{0}(980)$ mesons, which triggered the interest in other possible structures
for these mesons.

The classic study of tetraquark structures by Jaffe \cite{Jaffe} in the
framework of a quark-bag model predicted a light scalar nonet with an inverted
spectrum which nowadays can be identified with a nonet composed by the
$f_{0}(980)$, $a_{0}(980),$ $f_{0}(600)$ (or $\sigma$) and the more
controversial $\kappa(900)$. In this framework, the inverted mass spectrum and
couplings are\ related to the flavor structure. A strong attractive
gluon-magnetic interaction is essential here in order to pull the natural
scale $M\sim4m_{q}$ down to $M\approx900\ MeV$. Similar results for the
members of the scalar nonet are obtained in a variety of formalisms. In
particular, a recent analysis reach this conclusion analyzing three different
independent aspects of the problem \cite{Oller}. The extensive literature to
the scalar mesons problem can be traced back from this work.

Recently, chiral Lagrangians have been formulated trying to understand the
spectrum and interactions of scalar mesons. Clearly, the effective Lagrangian
approach would be able to describe the long distance physics of these mesons
in contrast with quark model descriptions which are based on the internal
degrees of freedom therefore taking into account effects at distances of the
order of meson radius and smaller. Although chiral Lagrangians deal with
effective degrees of freedom (d.o.f.), in principle, transformation properties
of these d.o.f. can be mapped directly to a quark content. This mapping must
be taken carefully as discussed in \cite{syracuse}.

An alternative explanation to the inverted spectrum of light scalars was
formulated based on the properties of the QCD vacuum \cite{Mauro,MS}. This
explanation relies on the long distance description of these mesons embraced
in effective Lagrangians which take into account the main properties of the
QCD vacuum, the spontaneous breaking of chiral symmetry and the breaking of
the $U_{A}(1)$ symmetry by non-perturbative effects. The role of the effective
six-quark interaction due to instantons \cite{thooft} in the inversion of the
light scalar spectrum was emphasized in \cite{MS}.

The existence of two scalar nonets, one below $1\ GeV$ and another one around
$1.4\ GeV$, lead to the exploration of two nonet models. A first step in this
direction was given in \cite{twononetsyr} where a light four-quark scalar
nonet and a heavy $\overline{q}q$ -like nonet are mixed up to yield physical
scalar mesons above and below $1$ $GeV$. In the same spirit, in Ref.
\cite{syracuse} a linear sigma model is coupled to a nonet field with well
defined transformation properties under chiral rotations and trivial dynamics,
except for a $U_{A}(1)$ (non-determinantal) violating interaction. The
possibility for the pseudoscalar light mesons to have a small four-quark
content was speculated on the light of this \textquotedblleft
toy\textquotedblright\ model. An alternative approach was formulated in
\cite{TC} where two linear sigma models were coupled using the very same
interaction as in \cite{twononetsyr} . The novelty here is a Higss mechanism
at the hadron level in such a way that the members of a pseudoscalar nonet are
\textquotedblleft eaten up\textquotedblright\ by axial vector mesons and one
ends up with only one nonet of pseudoscalars and two nonets of scalars. The
model is used as a specific realization of the conjectured \textquotedblleft
energy-dependent\textquotedblright\ composition of scalar mesons based on a
detailed analysis of the whole experimental situation \cite{TC}.

In the energy region around $1.4\ $GeV the Particle Data Group lists the
following scalar states : $f_{0}(1370),$ $K_{0}^{\ast}(1430),$ $a_{0}(1450),$
and $f_{0}(1500)$ \cite{PDG}.\ In addition we have the $f_{0}(1710)$ at a
slightly higher mass \cite{PDG}. We note also that if we consider the former
states as the members of a nonet then it presents a slightly inverted mass
spectrum. Furthermore, a look onto the pseudoscalar side at the same energy
yields the following states: $\eta(1295),$ $\eta(1440),$ $K(1460),$
$\pi(1300)$ \cite{PDG}. Thus \emph{data seems to indicate the existence of a
cuasi-degenerate chiral nonet around }$\emph{1.4}$\emph{ GeV} \emph{whose
scalar component has a slightly inverted mass spectrum}. On the other hand,
the linear rising of the mass of a hadron with the number of constituent
quarks indicates that four-quark states should lie slightly below $1.5$ GeV.
This lead us to conjecture that this chiral nonet comes from tetraquark states
mixed with conventional $\overline{q}q$ mesons to form physical mesons. The
cuasi-degeneracy of this chiral nonet suggest that chiral symmetry is realized
in a direct way for tetraquark states.

In this letter we report results on the implementation of this idea in the
framework of an effective chiral Lagrangian. We start with two chiral nonets,
one around $1.4$ $GeV$ with chiral symmetry realized directly and another one
at low energy with chiral symmetry spontaneously broken. In contrast to
previous studies, mesons in the \textquotedblleft heavy\textquotedblright%
\ nonet are considered as four-quark states. The nature of these states is
distinguished from conventional $\overline{q}q$ mesons by terms breaking the
$U_{A}(1)$ symmetry. We introduce also mass terms appropriate to four-quark
states which yield an inverted spectrum for the \textquotedblleft
pure\textquotedblright\ four-quark structured fields. These states mix with
conventional $\overline{q}q$ states to yield physical mesons. As a result we
obtain weak mixing in the isospinor and isovector pseudoscalar sectors hence
the $\pi(137)$ and $K(495)$ mesons arise as mainly $\overline{q}q$ states
whereas the $\pi(1300)$ and $K(1460)$ turn out to be mainly tetraquarks. In
contrast, scalars in these isotopic sectors are strongly mixed thus the
$a_{0}(980)$ $\kappa(900)$ states are strong admixtures of $\overline{q}q$ and
tetraquark states. Concerning isoscalar pseudoscalars we also find
$\eta(1440)$ and $\eta(958)$ as strong admixtures of $\overline{q}q$ and
tetraquarks whereas $\eta(547)$ and $\eta(1295)$ turn out to be mainly
$\overline{q}q$ and tetraquark states respectively. As for isoscalar scalar
mesons we find the $f_{0}(600)$ and $f_{0}(1370)$ as mainly $\overline{q}q$
and tetraquark states respectively whereas the $f_{0}(980)$ and $f_{0}(1500)$
arise as strong admixtures of $\overline{q}q$ and tetraquark states. We remark
that results for isosinglet scalars are expected to get modified by the mixing
of pure $\overline{q}q$ and four-quark- structured mesons with the lowest
lying scalar glueball field.

The idea of a chiral tetraquark nonet is implemented in the framework of an
effective model written in terms of \ \textquotedblleft
standard\textquotedblright\ ($\overline{q}q$-like ground states) meson fields
and \ \textquotedblleft non-standard\textquotedblright\ ( four-quark like
states) fields denoted as$\ \Phi=S+iP$ and $\hat{\Phi}=\hat{S}+i\hat{P}$
respectively. Here, $S$ , $\hat{S}$ and $P$ , $\hat{P}$ denote matrix fields
defined as%

\begin{equation}
F\equiv\frac{1}{\sqrt{2}}f_{i}\lambda_{i}\qquad i=1,...,7,ns,s
\label{notation}%
\end{equation}
where $f_{i}$\ stands for a generic field, $\lambda_{i}$ $(i=1,...,7)$ denote
Gell-Mann \ matrices and we work in the strange-non-strange basis for the
isoscalar sector, i.e. we use $\lambda_{ns}=diag(1,1,0)$ \ and $\lambda
_{s}=\sqrt{2}diag(0,0,1)$. Explicitly, the bare $S$ , $\hat{S}$ scalar and $P$
, $\hat{P}$ pseudoscalar nonets are given by%
\begin{equation}
S=\left(
\begin{array}
[c]{ccc}%
\frac{S_{ns}+S^{0}}{\sqrt{2}} & S^{+} & Y^{+}\\
S^{-} & \frac{S_{ns}-S^{0}}{\sqrt{2}} & Y^{0}\\
Y^{-} & \bar{Y}^{0} & S_{s}%
\end{array}
\right)  ;P=\left(
\begin{array}
[c]{ccc}%
\frac{H_{ns}+p^{0}}{\sqrt{2}} & p^{+} & X^{+}\\
p^{-} & \frac{H_{ns}-p^{0}}{\sqrt{2}} & X^{0}\\
X^{-} & \bar{X}^{0} & H_{s}%
\end{array}
\right)  , \label{qqfields}%
\end{equation}%
\begin{equation}
\hat{S}=\left(
\begin{array}
[c]{ccc}%
\frac{\hat{S}_{ns}+\hat{S}^{0}}{\sqrt{2}} & \hat{S}^{+} & \hat{Y}^{+}\\
\hat{S}^{-} & \frac{\hat{S}_{ns}-\hat{S}^{0}}{\sqrt{2}} & \hat{Y}^{0}\\
\hat{Y}^{-} & \overline{\hat{Y}}^{0} & \hat{S}_{s}%
\end{array}
\right)  ;\hat{P}=\left(
\begin{array}
[c]{ccc}%
\frac{\hat{H}_{ns}+\hat{p}^{0}}{\sqrt{2}} & \hat{p}^{+} & \hat{X}^{+}\\
\hat{p}^{-} & \frac{\hat{H}_{ns}-\hat{p}^{0}}{\sqrt{2}} & \hat{X}^{0}\\
\hat{X}^{-} & \overline{\hat{X}}^{0} & \hat{H}_{s}%
\end{array}
\right)  . \label{4qfields}%
\end{equation}

Next we implement the idea of a chiral nonet around 1.4 GeV using an effective
linear Lagrangian in terms of the four-quark structured fields $\hat{\Phi}$
with chiral symmetry realized linearly and directly ( $\hat{\mu}^{2}>0$)%

\begin{align}
\mathcal{L}(\hat{\Phi})  &  =\mathcal{L}_{sym}(\hat{\Phi})+\mathcal{L}%
_{SB}^{(1)}(\hat{\Phi})\label{L4q}\\
\mathcal{L}_{sym}(\hat{\Phi})  &  =\frac{1}{2}\left\langle \partial_{\mu}%
\hat{\Phi}\text{ }\partial^{\mu}\hat{\Phi}^{\dagger}\right\rangle -\frac
{\hat{\mu}^{2}}{2}\left\langle \hat{\Phi}\hat{\Phi}^{\dagger}\right\rangle
-\frac{\widehat{\lambda}}{4}\left\langle (\hat{\Phi}\hat{\Phi}^{\dagger}%
)^{2}\right\rangle \nonumber\\
\mathcal{L}_{SB}^{(1)}(\hat{\Phi})  &  =-\frac{\hat{c}}{4}\left\langle
\mathcal{M}_{Q}(\hat{\Phi}\hat{\Phi}^{\dagger}+\hat{\Phi}^{\dagger}\hat{\Phi
})\right\rangle \nonumber
\end{align}
Here, $\hat{\mu}$ sets the scale at which four-quark states lie, it is
expected to be slightly below 1.4 GeV. Although some of the fields in
$\hat{\Phi}$ have the same quantum numbers as the vacuum they do not acquire
vacuum expectation values (vev%
\'{}%
s) since we require a direct realization of chiral symmetry. The symmetry
breaking term in Eq. (\ref{L4q}) requires some explanation. This is an
explicit breaking term quadratic in the fields and proportional to a quadratic
quark mass matrix. The point is that in a quiral expansion the quark mass
matrix has a non-trivial flavor structure and enters as an external scalar
field. In the case of $\ $a four-quark nonet there must be breaking terms with
the appropriate flavor structure. This matrix is constructed according to the
flavor structure of the quark mass matrix as%

\begin{equation}
(\mathcal{M}_{Q})_{a}^{\ \ b}=\frac{1}{2}\epsilon_{acd}\epsilon^{bef}%
(\mathcal{M}_{q}{^{\dagger})}_{e}^{\ \ c}(\mathcal{M}_{q}{^{\dagger})}%
_{f}^{\ \ d}, \label{M4q}%
\end{equation}
where $\mathcal{M}_{q}=Diag(m,m,m_{s})$ stands for the conventional quark mass
matrix in the good isospin limit which we will consider henceforth. Explicitly
we obtain $\mathcal{M}_{Q}=Diag(mm_{s},mm_{s},m^{2})$. This structure yields
to pure 4q-structured fields an inverted spectrum with respect to conventional
states. A word of caution is necessary concerning the notation in Eq.
(\ref{4qfields}). The matrix field for four-quark states has a schematic quark
content%
\[
\hat{\Phi}\sim\left(
\begin{array}
[c]{ccc}%
\overline{q}q\overline{s}s & \overline{q}q\overline{s}s & \overline
{q}q\overline{q}s\\
\overline{q}q\overline{s}s & \overline{q}q\overline{s}s & \overline
{q}q\overline{q}s\\
\overline{q}q\overline{q}s & \overline{q}q\overline{q}s & \overline
{q}q\overline{q}q
\end{array}
\right)  ,
\]
where $q$ denote a $u$ or $d$ quark. The subindex $s$ and $ns$ in these fields
refer to the notation for $SU(3)$ matrices in Eq. (\ref{notation}) but do not
correspond with the hidden quark-antiquark content, e.g $\hat{S}_{ns}%
\sim\overline{q}q\overline{s}s$ and $\hat{S}_{s}\sim\overline{q}q\overline
{q}q$ .

Conventional $\overline{q}q$-structured fields are introduced in a chirally
symmetric way with chiral symmetry spontaneously broken. We introduce also an
instanton inspired breaking for the $U_{A}(1)$ symmetry. Notice that the
determinantal interaction is appropriate for $\overline{q}q$-structured fields
but not for four-quark fields since this is a six-quark interaction%

\begin{align}
\mathcal{L}(\Phi)  &  =\mathcal{L}_{sym}(\Phi)+\mathcal{L}_{A}+\mathcal{L}%
_{SB}^{(2)}(\Phi)\label{L2q}\\
\mathcal{L}_{sym}(\Phi)  &  =\frac{1}{2}\left\langle \partial_{\mu}%
\Phi\partial^{\mu}\Phi^{\dagger}\right\rangle -\frac{\mu^{2}}{2}\left\langle
\Phi\Phi^{\dagger}\right\rangle -\frac{\lambda}{4}\left\langle (\Phi
\Phi^{\dagger})^{2}\right\rangle \nonumber\\
\mathcal{L}_{A}  &  =-B\left(  \det\Phi+\det\Phi^{\dagger}\right)
,\quad\mathcal{L}_{SB}^{(2)}(\Phi)=\frac{b_{0}}{\sqrt{2}}\left\langle
\mathcal{M}_{q}\left(  \Phi+\Phi^{\dagger}\right)  \right\rangle \label{LSB1}%
\end{align}

This is just the model used in \cite{Mauro, MS} except for an OZI-forbidden
interaction whose corresponding coupling is consistent with zero when included
in the present context. The Lagrangian
\[
\mathcal{L}_{sym}(\Phi,\hat{\Phi})=\mathcal{L}_{sym}(\Phi)+\mathcal{L}%
_{sym}(\hat{\Phi})
\]
is invariant under the independent chiral transformations%

\begin{align}
\Phi &  \rightarrow U_{L}(\alpha_{L})\Phi U_{R}^{\dagger}(\alpha_{R}%
),\qquad\hat{\Phi}\rightarrow\hat{\Phi}\\
\Phi &  \rightarrow\Phi,\qquad\qquad\qquad\qquad\quad\hat{\Phi}\rightarrow
\hat{U}_{L}(\hat{\alpha}_{L})\hat{\Phi}\hat{U}_{R}^{\dagger}(\hat{\alpha}%
_{R}).
\end{align}
i.e. it has $\left(  U(3)\times U(3)\right)  ^{2}$ symmetry. This symmetry is
explicitly broken down to $SU(3)_{A}\times U(3)_{V}$ by the interaction%
\[
\mathcal{L}_{\epsilon^{2}}=-\frac{\epsilon^{2}}{4}\left\langle \Phi\hat{\Phi
}^{\dagger}+\hat{\Phi}\Phi^{\dagger}\right\rangle .
\]
Further sources of breaking come from the anomaly term and quark mass terms in
Eqs. (\ref{L4q}, \ref{L2q}). Finally, since we are considering quark masses
entering as external scalar fields we also consider the following terms%

\begin{equation}
\mathcal{L}_{SB}^{(3)}=\frac{\hat{b}_{0}}{\sqrt{2}}\left\langle \mathcal{M}%
_{q}(\hat{\Phi}+\hat{\Phi}^{\dagger})\right\rangle +\frac{\hat{d}}{\sqrt{2}%
}\left\langle \mathcal{M}_{Q}(\hat{\Phi}+\hat{\Phi}^{\dagger})\right\rangle .
\label{Lsb3}%
\end{equation}

A last term proportional to $\left\langle \mathcal{M}_{Q}(\Phi+\Phi^{\dagger
})\right\rangle $ can also be added without altering the conclusions of this
work. The linear terms in (\ref{LSB1}) induce scalar-to-vacuum transitions
which instabilizes the vacuum. We shift to the true vacuum, $S\rightarrow S-V$
where $V$ stands for the vacuum expectation values of $S$ which we denote as
$V=diag(a,a,b)$. This mechanisms generates new meson mass terms and
interactions. In particular, the shift generates linear terms in $\hat{\Phi}$
which cancel against the linear terms in Eq.(\ref{Lsb3}). Here, we present
results for meson masses, details of the calculations and results for
interactions will be published elsewhere \cite{prep}.\ 

For the isodoublets and isotriplets,\ the interaction term $\mathcal{L}%
_{\epsilon^{2}}$ mix up $\overline{q}q$ and four-quark states. We define the
diagonal isotriplet pseudoscalar fields as
\[
\left(
\begin{array}
[c]{c}%
\pi\\
\hat{\pi}%
\end{array}
\right)  =\left(
\begin{array}
[c]{cc}%
\cos(\theta_{1}) & -\sin(\theta_{1})\\
\sin(\theta_{1}) & \cos(\theta_{1})
\end{array}
\right)  \left(
\begin{array}
[c]{c}%
p\\
\hat{p}%
\end{array}
\right)  .
\]

For the isotriplet scalar sector we denote the physical fields as $a$ , $A$
and the corresponding mixing angle is denoted by $\phi_{1}$. For the
isodoublets we denote the physical fields as $K,\hat{K}$ ($\kappa,\hat{\kappa
})$, and the mixing angle as $\theta_{1/2}$ ( $\phi_{1/2})$ for pseudoscalars (scalars).

The isosinglet sectors are more involved due to the effects coming from the
$U_{A}(1)$ anomaly which when combined with the spontaneous breaking of chiral
symmetry produces mixing among four different fields. The mass Lagrangian for
the isoscalar pseudoscalar sector reads%
\begin{equation}
\mathcal{L}_{H}=-\frac{1}{2}\left\langle H\right\vert M_{H}\left\vert
H\right\rangle
\end{equation}
where%
\begin{equation}
\left\vert H\right\rangle =\left(
\begin{array}
[c]{c}%
H_{ns}\\
H_{s}\\
\hat{H}_{ns}\\
\hat{H}_{s}%
\end{array}
\right)  ,\quad M_{H}=\left(
\begin{array}
[c]{cccc}%
m_{H_{ns}}^{2} & m_{H_{s-ns}}^{2} & \frac{\epsilon^{2}}{2} & 0\\
m_{H_{s-ns}}^{2} & m_{H_{s}}^{2} & 0 & \frac{\epsilon^{2}}{2}\\
\frac{\epsilon^{2}}{2} & 0 & m_{\hat{H}_{ns}}^{2} & 0\\
0 & \frac{\epsilon^{2}}{2} & 0 & m_{\hat{H}_{s}}^{2}%
\end{array}
\right)  , \label{MH}%
\end{equation}
\newline with%

\[%
\begin{array}
[c]{ll}%
\ \ m_{H_{ns}}^{2}=\mu^{2}-2Bb+\lambda a^{2}, & m_{\hat{H}_{ns}}^{2}=\hat{\mu
}^{2}+\hat{c}mm_{s},\\
\ \ m_{H_{s}}^{2}\ =\mu^{2}+\lambda b^{2}, & m_{\hat{H}_{s}}^{2}\ =\hat{\mu
}^{2}+\hat{c}m^{2}.\\
m_{H_{s-ns}}^{2}=-2\sqrt{2}Ba, &
\end{array}
\]

For the isoscalar-scalar sector we obtain a similar structure. In principle,
these matrices are diagonalized by a general rotation in $O(4)$ containing six
independent parameters. However, modulo corrections of the order
$m(m_{s}-m)/\hat{\mu}^{2}$, it can be shown that they are actually
diagonalized by a rotation in the subgroup $SO(2)\otimes SO(2)\otimes SO(2)$
and a straightforward analytic solution depending on three angles is obtained
for the rotation matrix. Explicitly, under this approximation the physical
pseudoscalar fields are given as
\[
\left(
\begin{array}
[c]{c}%
\eta\\
\eta^{\prime}\\
\hat{\eta}\\
\hat{\eta}^{\prime}%
\end{array}
\right)  =\left(
\begin{array}
[c]{cccc}%
c_{\alpha}c_{\beta} & -s_{\alpha}c_{\beta} & -c_{\alpha}s_{\beta} & s_{\alpha
}s_{\beta}\\
s_{\alpha}c_{\beta^{\prime}} & c_{\alpha}c_{\beta^{\prime}} & s_{\alpha
}s_{\beta^{\prime}} & c_{\alpha}s_{\beta^{\prime}}\\
c_{\alpha}s_{\beta} & -s_{\alpha}s_{\beta} & c_{\alpha}c_{\beta} & -s_{\alpha
}c_{\beta}\\
-s_{\alpha}s_{\beta}^{\prime} & -c_{\alpha}c_{\beta^{\prime}} & s_{\alpha
}c_{\beta^{\prime}} & c_{\alpha}c_{\beta^{\prime}}%
\end{array}
\right)  \left(
\begin{array}
[c]{c}%
{\small H}_{ns}\\
{\small H}_{s}\\
{\small \hat{H}}_{ns}\\
{\small \hat{H}}_{s}%
\end{array}
\right)
\]
where $c_{\alpha}\equiv\cos\alpha$ , $s_{\alpha}\equiv\sin\alpha$. Similar
relations are also valid for the scalar sector. We denote as $\sigma,$ $f_{0}%
$, $\hat{\sigma},$ $\hat{f}_{0}$ to the physical isosinglet scalar fields and
$\gamma,$ $\delta$ and $\delta^{\prime}$ to the mixing angles analogous to
$\alpha,$ $\beta,$ $\beta^{\prime}$ respectively.

Finally, a calculation of the axial currents allow us to relate the vacuum
expectation values of scalars to the weak decay constants of pseudoscalars as%
\begin{equation}
a=\frac{f_{\pi}}{\sqrt{2}\cos\theta_{1}},\ \quad a+b=\frac{\sqrt{2}f_{K}}%
{\cos\theta_{1/2}}. \label{amb1}%
\end{equation}

\begin{figure}
[ptb]
\begin{center}
\includegraphics[
height=2.4749in,
width=3.9095in
]%
{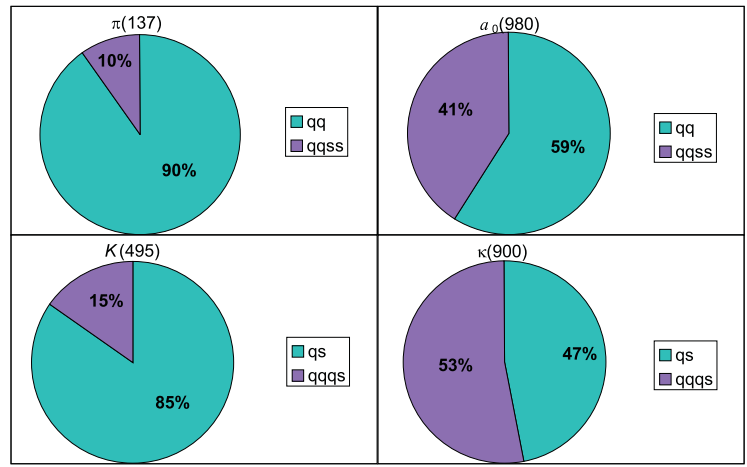}%
\caption{Quark content of light isotriplet and isodoublets.}%
\label{fig1}%
\end{center}
\end{figure}

\begin{figure}
[ptb]
\begin{center}
\includegraphics[
height=2.8833in,
width=4.5512in
]%
{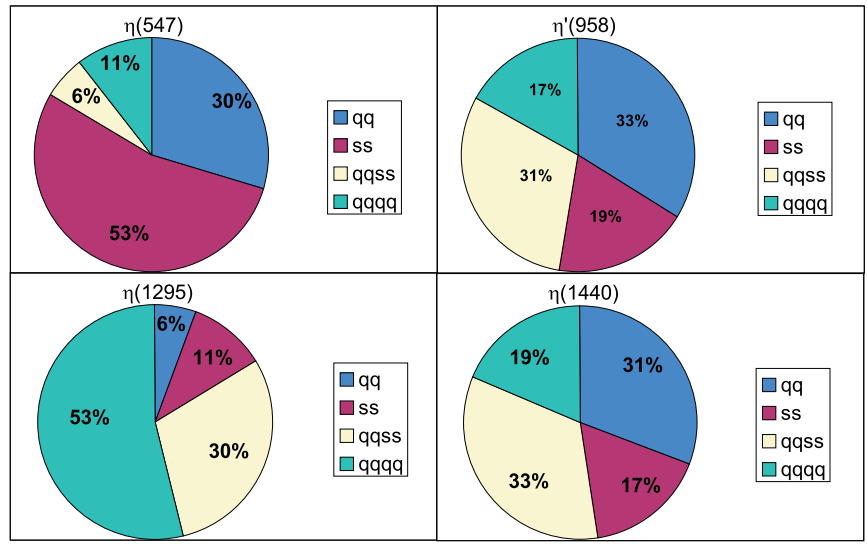}%
\caption{Quark content of light isosinglet pseudoscalars.}%
\label{fig2}%
\end{center}
\end{figure}

\begin{figure}
[ptb]
\begin{center}
\includegraphics[
height=2.8808in,
width=4.5512in
]%
{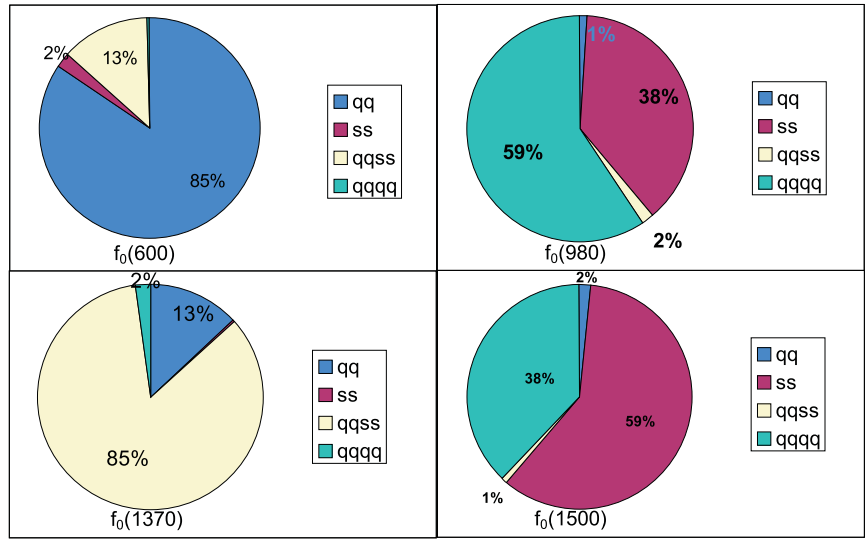}%
\caption{Quark content of light isosinglet scalars.}%
\label{fig3}%
\end{center}
\end{figure}

There are eight free parameters in the model which are relevant to meson
masses: $\{\mu^{2},$ $\lambda,$ $B,$ $\epsilon^{2},$ $a,$ $b,$ $\hat{\mu}%
_{1}^{2},$ $\hat{\mu}_{1/2}^{2}\}$, where $\hat{\mu}_{1/2}^{2}\equiv\hat{\mu
}^{2}+\frac{\widehat{c}}{2}m(m+m_{s})$ and $\hat{\mu}_{1}^{2}\equiv\hat{\mu
}^{2}+\widehat{c}mm_{s}$.\ Our input are the physical quantities listed in
Table I , namely, the masses for $\pi(137)$ , $a_{0}(980)$ , $K(495),$
$\eta(547),$ $\eta^{\prime}(958),$ $\eta(1295)$ in addition to the weak decay
constants $f_{\pi}$ and $f_{K}$. Uncertainties listed in this table correspond
to the measured values in the case of the isosinglets \cite{PDG}. Since we are
working in the good isospin limit we use the experimental deviations from this
limit for the uncertainties in the masses of isotriplets and isodoublets.
Using these values we fix the parameters to the values also listed in Table I.
In Table II we show the predictions of the model for the remaining meson
masses and mixing angles and the experimental values for these quantities when
available. The quark content of mesons corresponding to the central values of
these mixing angles are shown in Figs. \ref{fig1}, \ref{fig2}, \ref{fig3}. In
Fig. 1 we show results for the light isodoublets and isotriples. Heavy mesons
have the opposite quark content. We obtain pions and kaons as mainly
$q\overline{q}$ states whereas the heavy fields $\pi(1300),$ $K(1460)$ arise
as mainly tetraquark states. In contrast, in the scalar sector isotriplets and
isodoublets get strongly mixed and the physical mesons turn out to have almost
identical amounts of $q\overline{q}$ and four-quark content. In Fig. 2 we show
results for isosinglet pseudoscalars. Here, in the case of $\eta(547)$ we
obtain also a small four-quark content, the $\eta(1295)$ being almost a
four-quark state. However, the $\eta^{\prime}(958)$ and $\eta(1440),$ turn out
to be strong admixtures of $q\overline{q}$ and four-quark states with almost
equal amounts of them. As to the isosinglet scalars Fig. 3 shows that the
sigma meson ( $f_{0}(600)$) arises as mainly $q\overline{q}$ state and the
opposite quark content is carried by the $f_{0}(1370)$ which turns out to be
mainly a four quark state. Finally, the $f_{0}(980)$ turns out to be have
roughly a 40\% content of $\overline{q}q$ (more explicitly $\overline{s}s$ )
and 60\% of $\overline{q}\overline{q}qq$ whereas the $f_{0}(1500)$ is composed
60\% of $\overline{q}q$ (more explicitly $\overline{s}s$ ) and almost 40\% of
$\overline{q}\overline{q}qq$. We stress again that quark content of isoscalar
scalar mesons shown in Fig. 3 are expected to be modified by the inclusion of
the scalar glueball in the present context although we do not expect strong
modifications in the case of the $f_{0}(600)$ which is far from the energy
region around 1.6 GeV where this state is expected.

Summarizing, in this work we point out the possibility that scalar and
pseudoscalar mesons below $1.5$ GeV be admixtures of conventional
$\overline{q}q$ and tetraquark states. This conjecture is risen by the key
observation that beyond the light scalars lying below $1\ GeV$, there are nine
states around 1.4GeV which, when considered as the members of a nonet, exhibit
a slightly inverted mass spectrum as compared with a conventional
$\overline{q}q$ nonet. Furthermore this nonet lie at an energy scale
compatible with that of tetraquarks according to the linear rising of the mass
of a hadron with the number of constituents. There is a nonet of pseudoscalar
states at the same mass which suggest these nonets form a chiral nonet. We
implement this idea in the framework of a chiral model. As a result we obtain
a meson spectrum consistent with the measured pseudoscalar and scalar meson
spectrum below 1.5 GeV.

\begin{acknowledgments}
Work supported by CONACyT, M\'{e}xico under project 37234-E.\bigskip
\end{acknowledgments}

\newpage

\begin{center}

Tabe I\bigskip%

\begin{tabular}
[c]{||l||l||l||l||}\hline\hline
& Input & Parameter & Fit\\\hline\hline
$m_{\pi}$ & $137.3\pm2.3\ $MeV & $\hat{\mu}_{1}(${\small MeV}$)$ &
$1257.5\pm61.3$\\\hline\hline
$m_{a}$ & $984.3\pm0.9\ $MeV & $\hat{\mu}_{1/2}(${\small MeV}$)$ &
$1206.2\pm142.3$\\\hline\hline
$f_{\pi}$ & $92.42\pm3.53\ $MeV & $\left\vert \epsilon\right\vert
(${\small MeV}$)$ & $1012.3\pm75.9$\\\hline\hline
$m_{K}$ & $495.67\pm2.00\ $MeV & $a(${\small MeV}$)$ & $68.78\pm
4.47$\\\hline\hline
$f_{K}$ & $113.0\pm1.3\ $MeV & $b(${\small MeV}$)$ & $104.8\pm2.6$%
\\\hline\hline
$m_{\eta}$ & $547.30\pm0.12\ $MeV & $B(${\small GeV}$)$ & $-2.16\pm
0.28$\\\hline\hline
$m_{\eta^{\prime}}$ & $957.78\pm0.14\ $MeV & $\lambda$ & $31.8\pm
5.9$\\\hline\hline
$m_{\hat{\eta}}$ & $1297.0\pm2.8\ $MeV & $\mu^{2}(${\small GeV}$^{2})$ &
$0.490\pm0.107$\\\hline\hline
\end{tabular}

Table I \ \ Input used to fix the parameters entering the model and their values.

$\bigskip$

Table II$\bigskip$

$%
\begin{tabular}
[c]{|l|l|l|l|}\hline
Mass & Prediction & Identification & Exp\\\hline
$m_{\hat{\pi}}($MeV$)$ & $1322.6\pm32.4$ & $\pi(1300)$ & $1300\pm
200$\cite{PDG}\\\hline
$m_{\hat{K}}($MeV$)$ & $1293.1\pm3.5$ & $K(1460)$ & $1400-1460$\cite{PDG}%
\\\hline
$m_{A}($MeV$)$ & $1417.3\pm51.0$ & $a_{0}(1450)$ & $1474\pm19$\cite{PDG}%
\\\hline
$m_{\kappa}($MeV$)$ & $986.2\pm19.1$ & $\kappa(900)$ & $750-950$%
\cite{E791,Svec,Black:1998zc}\\\hline
$m_{\hat{\kappa}}($MeV$)$ & $1413.9\pm76.4$ & $K_{0}^{\ast}(1430)$ &
$1429\pm4\pm5$\cite{PDG}\\\hline
$m_{\hat{\eta}}^{\prime}($MeV$)$ & $1394.0\pm61.9$ & $\eta(1440)$ &
$1400-1470$\cite{PDG}\\\hline
$m_{\sigma}($MeV$)$ & $380.6\pm91.0$ & $f_{0}(600)$or $\sigma$ & $478$ $\pm$
$35$\cite{E791}\\\hline
$m_{f_{0}}($MeV$)$ & $1022.4\pm25.6$ & $f_{0}(980)$ & $980\pm10$%
\cite{PDG}\\\hline
$m_{\hat{\sigma}}($MeV$)$ & $1284.7\pm15.3$ & $f_{0}(1370)$ & $1200-1500$%
\cite{PDG}\\\hline
$m_{\hat{f}_{0}}($MeV$)$ & $1447.7\pm84.6$ & $f_{0}(1500)$ & $1500\pm
10$\cite{PDG}\\\hline
$\theta_{1}$ & $18.16^{\circ}\pm4.34^{\circ}$ &  & \\\hline
$\theta_{1/2}$ & $22.96^{\circ}\pm4.84^{\circ}$ &  & \\\hline
$\phi_{1}$ & $39.8^{\circ}\pm4.5^{\circ}$ &  & \\\hline
$\phi_{1/2}$ & $46.7^{\circ}\pm9.5^{\circ}$ &  & \\\hline
$\alpha$ & $53.4^{\circ}\pm0.8^{\circ}$ &  & \\\hline
$\beta$ & $23.9^{\circ}\pm5.1^{\circ}$ &  & \\\hline
$\beta^{\prime}$ & $43.6^{\circ}\pm7.2^{\circ}$ &  & \\\hline
$\gamma$ & $-9.11^{\circ}\pm0.49^{\circ}$ &  & \\\hline
$\delta$ & $21.45^{\circ}\pm6.49^{\circ}$ &  & \\\hline
$\delta^{\prime}$ & $51.36^{\circ}\pm8.35^{\circ}$ &  & \\\hline
\end{tabular}
\ \ \ \ \ $\bigskip

Table II \ Predictions of the model for meson masses and mixing angles.
\end{center}

%

%

%

\end{document}